\documentclass[a4paper]{article}

\usepackage{a4wide}
\usepackage{hyperref}
\usepackage{graphicx}
\usepackage{array}
\usepackage{amsmath}
\usepackage{dsfont}
\usepackage[plain,vlined]{algorithm2e}
\def\T{{ \mathrm{\scriptscriptstyle T} }}

\title{Variational Bayes for Gaussian Factor Models under the Cumulative Shrinkage Process}
\author{Sirio Legramanti\\ \small Department of Decision Sciences, Bocconi University, 20136 Milan, Italy\\ \small \texttt{sirio.legramanti@unibocconi.it}}
\date{}

\newtheorem{theorem}{Theorem}[section]
\newtheorem{definition}[theorem]{Definition}

\begin{document}
	\maketitle
	
\begin{abstract}
	The cumulative shrinkage process is an increasing shrinkage prior that can be employed within models in which additional terms are supposed to play a progressively negligible role.
	A natural application is to Gaussian factor models, where such a process has proved effective in inducing parsimonious representations while providing accurate inference on the data covariance matrix. 
	The cumulative shrinkage process came with an adaptive Gibbs sampler that tunes the number of latent factors throughout iterations, which makes it faster than the non-adaptive Gibbs sampler.
	In this work we propose a variational algorithm for Gaussian factor models endowed with a cumulative shrinkage process.
	Such a strategy provides comparable inference with respect to the adaptive Gibbs sampler and further reduces runtime.\\
	
\end{abstract}

\section{Introduction}
\label{sec_intro}

The cumulative shrinkage process \cite{legramanti20cusp} is an increasing shrinkage prior based on a sequence of spike-and-slab distributions, with growing mass assigned to the spike. 
It can be defined for both countable and finite sequences.
A definition for the countable case can be found in \cite{legramanti20cusp}, while the following is a definition for finite sequences.

\begin{definition} \label{def_cusp}
	We say that 
	$\{ \theta_h \in \Theta \subseteq \mathds{R}: h=1,\ldots,H \}$ is distributed according to a cumulative shrinkage process with shrinkage parameter $ \alpha>0 $, slab $P_0$ and spike $P_\infty$ if, conditionally on $\{ \pi_h \in (0,1): h=1,\ldots,H \} $, each $ \theta_h $ is independent and
	\begin{equation} \label{eq_cusp1}
	(\theta_{h} \mid \pi_h) \sim 
	( 1 - \pi_h )  P_0 + \pi_h  P_\infty, \qquad (h=1,\ldots,H),
	\end{equation}
	where $ \pi_h = \sum\nolimits_{l=1}^{h} \omega_l$  for $ h=1,\ldots,H $ and 
	$ \omega_l= v_l \prod\nolimits_{m=1}^{l-1} (1-v_m) $ for $ l=1,\ldots,H $, with $ v_1, \ldots, v_{H-1} $ being independent  $\mbox{\normalfont Beta}(1,\alpha)$ random variables and $ v_H = 1 $.
\end{definition}
This construction, based on the stick-breaking representation of the Dirichlet process \cite{ishwaranjames}, implies that the sequence $ \pi_h $ is non-decreasing and that $ \pi_H=1 $.

The cumulative shrinkage process can be used in a variety of models, e.g. Poisson factorization
\cite{dunsonherring}, but here we focus on Gaussian factor models, which are ubiquitous in statistics and have been used in \cite{legramanti20cusp} as illustrative example.  
In~\cite{legramanti20cusp} posterior inference for this model under the cumulative shrinkage process is carried out through an adaptive Gibbs sampler which tunes $ H $ as it progresses.
This algorithm, together with the ability of the prior to favor the recovery of the number of active latent factors, allows for reduced runtime with respect to the non-adaptive Gibbs sampler.
However, the increasing availability of large datasets demands for even faster algorithms. 
This need for scalability has pushed Bayesian statisticians towards approximate methods for posterior inference, including Laplace approximation, variational Bayes and expectation propagation \cite{bishop2006pattern}. 

In this work we employ mean-field variational Bayes, which is straightforward to derive for Gaussian factor models under a convenient specification of the cumulative shrinkage process which slightly differs from the one in \cite{legramanti20cusp}.
Such a specification is detailed in \S~\ref{sec_model}, while the variational approximation is described in \S~\ref{sec_variational}. Finally, in \S~\ref{sec_application} we illustrate the performance of the variational algorithm on real data.

\section{Model and Prior}
\label{sec_model}

We focus on learning the structure of the $p \times p$ covariance matrix $\Omega=\Lambda \Lambda^{\T}+ \Sigma$ of the data $ y_i \in \mathds{R}^p $ from the Gaussian factor model $ {y_{i}=\Lambda \eta_i+\epsilon_i} $ $(i=1, \ldots, n)$, 
where $ \Lambda=[\lambda_{jh}] \in \mathds{R}^{p \times H} $, $\eta_{i} \sim N_H(0,I_H)$, $\epsilon_i \sim N_p(0, \Sigma)$ and $ {\Sigma = \mbox{diag}(\sigma_1^2,\ldots,\sigma_p^2)}$.
As priors, we let $\sigma^2_{j} \sim \mbox{InvGa}(a_{\sigma},b_{\sigma})$ for $j=1, \ldots, p$ and, differently from \cite{legramanti20cusp},
we place a cumulative shrinkage process directly on 
the loadings, with $ \pi_h $ as in Def.~\ref{def_cusp}:
\begin{equation} \label{eq_marginal}
(\lambda_{jh} \mid \pi_h) \sim ( 1 - \pi_h ) N(0,\theta_0) + 
\pi_h N(0,\theta_\infty),
\qquad (j=1,\ldots,p; \ h=1,\ldots,H).
\end{equation} 
This simpler specification facilitates the derivation of the variational algorithm, while preserving the increasing shrinkage property. In fact, setting $ \theta_0 > \theta_\infty $, the loadings are increasingly shrunk towards zero in probability, i.e. 
$ {\mbox{pr}\{ |\lambda_{j,h+1}|<\epsilon \}} \geq {\mbox{pr}\{ |\lambda_{jh}|<\epsilon \}} $ for any $ \epsilon>0 $, encoding the prior assumption that additional factors provide a decreasing contribution to the model.
However, setting both the spike and the slab to Gaussians is suboptimal to the specification in \cite{legramanti20cusp}, where the Student-t slab is more differentiated from the Gaussian spike, thus facilitating the separation of active and inactive factors.

The derivation of the variational algorithm is further facilitated by the introduction of the augmented data 
$ z_h = (z_{h1},\ldots,z_{hH}) \sim \mbox{Mult}\{1,(\omega_1,\ldots,\omega_H)\}$, which exploits the fact that equation~\eqref{eq_marginal} can be obtained by marginalizing out $ z_h $ from 
\begin{equation*} \label{eq_augm}
( \lambda_{jh} \mid z_h ) \sim 
\{1 - \sum\nolimits_{l=1}^{h} z_{hl}\} N(0,\theta_0)
+ \sum\nolimits_{l=1}^{h} z_{hl} N(0,\theta_\infty),  \quad (j=1,\ldots,p; \ h=1,\ldots,H).
\end{equation*}

\section{Variational Inference}
\label{sec_variational}

Variational Bayes approximates the posterior density with the density $ q^* $ that is closest to it, in Kullback-Leibler (KL) divergence, within a family $ Q $ of tractable densities (see \cite{blei2017} for a review). The ideal variational family $ Q $ should combine flexibility, that allows for a good approximation, and tractability.
Here we use the mean-field variational family, whose elements factorize as follows:
\begin{equation}\label{eq_variational_family}
q(\lambda,\eta,\sigma,z,v) = q(\lambda) q(\eta) q(\sigma) q(z) q(v).
\end{equation}
The KL divergence between such a $ q $ and the intractable posterior cannot be computed or minimized directly.
Equivalently, we maximize the evidence lower bound \begin{eqnarray} 
ELBO(q) &=& \log p(y) - KL(q(\lambda,\eta,\sigma,z,v)|| p(\lambda,\eta,\sigma,z,v \mid y))= \label{eq_elbo1}\\
&=& E_q[\log p(y,\lambda,\eta,\sigma,z,v)]-E_q[\log q(\lambda,\eta,\sigma,z,v)]. \label{eq_elbo2}
\end{eqnarray}
Equation \eqref{eq_elbo1} highlights that, since the KL divergence is always non-negative, the ELBO lower-bounds the log-evidence, thus justifying its name. 
Moreover, since $ \log p(y) $ does not depend on $ q $, maximizing the ELBO is equivalent to minimizing the KL divergence with respect to $ q $.
Since \eqref{eq_elbo1} involves the intractable posterior, the equivalent expression \eqref{eq_elbo2} is used to actually compute the ELBO.
The optimization is solved through coordinate ascent, iteratively maximizing the ELBO with respect to each factor on the right-hand side of~\eqref{eq_variational_family}. Following \cite[Ch. 10]{bishop2006pattern}, each factor update is derived as follows (we report only the loadings term for illustrative purposes):
\begin{equation*}
\log q^*(\lambda) = E_{\neq \lambda} [\log p(y,\lambda,\eta,\sigma,z,v)] + const,
\end{equation*}
where $ E_{\neq \lambda} $ denotes the expectation under $ q $ with respect to all variables other than the loadings.
With no parametric assumption on 
the factors in \eqref{eq_variational_family}, we obtain:
\begin{eqnarray*}
	q^*(\lambda,\eta,\sigma,z,v) &=& 
	\prod_{j=1}^{p} N_H(\lambda_{j \cdot}; \mu_j^{(\lambda)}, V_j^{(\lambda)})
	\prod_{i=1}^{n} N_H(\eta_{i \cdot}; \mu_i^{(\eta)}, V^{(\eta)})
	\prod_{j=1}^{p} \mbox{InvGa}(\sigma_j^2; A^{(\sigma)}, B_j^{(\sigma)}) \cdot\\
	&\cdot& \prod_{h=1}^{H} \mbox{Mult}(z_h; 1, \kappa_h)
	\prod_{h=1}^{H-1} \mbox{Beta}(v_h; A_h^{(v)}, B_h^{(v)}).\\
\end{eqnarray*}
Notice that each factor further factorizes into exponential-family distributions, thus facilitating computations. 
The update equations for the parameters are coupled, meaning that each factor update involves expectations with respect to other factors. 
We then proceed iteratively cycling over the steps of Algorithm~\ref{alg_variational}.
This routine converges to a local maximum, hence should be run from several initializations \cite{blei2017}.
Convergence of each run can be assessed by monitoring the monotone growth of the ELBO.
From the optimal variational parameters we can also compute the variational expectation of the number $ H^* $ of factors that are active, in the sense that they are modeled by the slab: 
$ E_{q^*}[H^*] = \sum\nolimits_{h=1}^{H} \sum\nolimits_{l=h+1}^{H} \kappa_{hl} $.

\begin{algorithm}[t]
	\caption{One cycle of the variational algorithm for Gaussian factor models} 
	\label{alg_variational}
	\nl \For{j from 1 to p}{
		set $ V_j^{(\lambda)} = \{ diag(\theta^*_1,\ldots,\theta^*_H) + ( A^{(\sigma)} / B_j^{(\sigma)} ) ( {\mu^{(\eta)}}^\T \mu^{(\eta)} + nV^{(\eta)} ) 
		\}^{-1}$,\\
		where $ \theta^*_h = (1-\sum_{l=1}^h \kappa_{hl})\theta_0^{-1}+(\sum_{l=1}^h \kappa_{hl})\theta_\infty^{-1} $,	and
		$ \mu_j^{(\lambda)} = ( A^{(\sigma)} / B_j^{(\sigma)} ) V_j^{(\lambda)} {\mu^{(\eta)}}^\T y_{\cdot j}$;
	}
	\nl Set $ A^{(\sigma)}=a_\sigma+n/2 $ and 
	\For{j from 1 to p}{ 
		$ { \mbox{set } {B_j^{(\sigma)} = b_\sigma + 0.5 \cdot \sum_{i=1}^{n} \{ y_{ij}^2 - 2 y_{ij} {\mu_i^{(\eta)}}^\T \mu_j^{(\lambda)} + \sum_{h=1}^{H} \sum_{k=1}^{H}(\mu_{ih}^{(\eta)}\mu_{ik}^{(\eta)}+V_{hk}^{(\eta)})
				(\mu_{jh}^{(\lambda)}\mu_{jk}^{(\lambda)}+V_{j;hk}^{(\lambda)})
				\};} }$
	}
	\nl Set $ V^{(\eta)}= (I_H + {\mu^{(\lambda)}}^\T diag( A^{(\sigma)} / B^{(\sigma)} ) \mu^{(\lambda)} + \sum_{j=1}^{p} ( A^{(\sigma)} / B_j^{(\sigma)} ) V_j^{(\lambda)}
	)^{-1}$;\\
	\For{i from 1 to n}{
		set $ \mu_i^{(\eta)}= V^{(\eta)} {\mu^{(\lambda)}}^\T diag( A^{(\sigma)} / B^{(\sigma)} ) y_{i \cdot}$;
	}
	\nl \For{h from 1 to H}{
		\lFor{l from 1 to h}{
			$ \mbox{set } \kappa_{hl} \propto$  $ {\exp \{ E(\log \omega_l) - 0.5 \cdot p \log \theta_\infty - 0.5 \cdot \theta_\infty^{-1} E[\lambda_{\cdot h}^T \lambda_{\cdot h}]
				\}} $}
		\lFor{l from h+1 to H}{			
			$ \mbox{set } \kappa_{hl} \propto$  $ {\exp \{ E(\log \omega_l) - 0.5 \cdot p \log \theta_0 - 0.5 \cdot \theta_0^{-1} E[\lambda_{\cdot h}^T \lambda_{\cdot h}]
				\}} $}
	}
	where $ E[\lambda_{\cdot h}^T \lambda_{\cdot h}] = \sum_{j=1}^{p} ( {\mu_{jh}^{(\lambda)}}^2 + V_{j;hh}^{(\lambda)} ) $ and, with $ \Psi $ being the digamma function, $ {E(\log \omega_l) = \mathds{1}\{l<H\} \{\Psi(A_l^{(v)})-\Psi(A_l^{(v)}+B_l^{(v)})\} 
		+ \mathds{1}\{l>1\} \sum_{m=1}^{l-1} \{\Psi(B_m^{(v)})-\Psi(A_m^{(v)}+B_m^{(v)})\}}$;\\
	\nl \For{h from 1 to $ (H-1) $}{
		set $ A_h^{(v)} = 1 + \sum_{l=1}^{H} \kappa_{lh} $ and $ B_h^{(v)} = \alpha + \sum_{l=1}^{H} \sum_{m=h+1}^H \kappa_{lm} $. 
	}
	\vspace{5pt}
\end{algorithm}

\section{Application to Personality Data}
\label{sec_application}

We compare our variational algorithm for the model in \S~\ref{sec_model} to the adaptive Gibbs sampler for the model proposed in~\cite{legramanti20cusp}, on the same real dataset considered there.
Namely, we consider a subset of the dataset \texttt{bfi} from the \texttt{R} package \texttt{psych}, containing the six-point-scale answers of $n=126$ individuals older than fifty years to $p=25$ 
questions about five personality traits.
As in~\cite{legramanti20cusp}, we center the 25 items and, to have coherent answers within each personality trait, we change sign to answers $1,9,10,11,12,22$ and $25$, as suggested in the documentation of the \texttt{bfi} dataset. 

For the adaptive Gibbs sampler, the model and the hyperparameters are specified as in \cite{legramanti20cusp}. 
For our variational algorithm, we set $ \alpha=5 $, $ \theta_0=1 $, $ {\theta_\infty = 10^{-6}} $ and we conservatively let $ H=p+1 $, which coincides with the initial value of $ H $ for the adaptive Gibbs sampler and corresponds to at most $ p $ latent factors.

We run the variational algorithm from 20 random initializations, stopping each run when the ELBO grows less than 0.05. We then pick the run reaching the highest ELBO. Using the optimal variational parameters of this run, we get a sample of size 2000 for $ \Omega $, from which we derive a sample for the correlation matrix $\Omega^*
=(\Omega \odot I_p)^{-1/2} \Omega (\Omega \odot I_p)^{-1/2}$, with $\odot$ denoting the element-wise product. 
From this sample we compute a Monte Carlo estimate of the mean squared deviations $\sum_{j=1}^p\sum_{q=j}^pE(\Omega^*_{jq}-S_{jq})^2/ \{p(p+1)/2 \}$ between $ \Omega^* $ and the sample correlation matrix~$S$. 
The same quantity is computed from a posterior sample of equal size obtained running the adaptive Gibbs sampler in~\cite{legramanti20cusp} for 10000 iterations after a burn-in of 5000 and then thinning every five.
The two quantities are reported as MSE (Mean Square Error) in Table~\ref{table_psych}, together with the expected number of active factors and the total running time for each of the two methods.

With respect to the adaptive Gibbs sampler, the proposed variational algorithm provides the same MSE (rounded off to the second decimal digit) and a similar expected number of active factors, but is more than five times faster.
\begin{table}[t]
	\caption{Performance of adaptive Gibbs sampler and variational algorithm on the \texttt{bfi} dataset}
	\label{table_psych}
	\begin{tabular}
		{p{0.3\textwidth}>{\centering}p{0.18\textwidth}>{\centering}p{0.2\textwidth}>{\centering\arraybackslash}p{0.2\textwidth}}
		\noalign{\smallskip}
		\hline
		\noalign{\smallskip}
		Method & MSE & E[H*] & Running time (s) \\
		\hline
		\noalign{\smallskip}
		Adaptive Gibbs sampler & 0.01 & 2.7 & 340 \\ 
		Variational inference & 0.01 & 3.0 & 63 \\ 
		\hline
	\end{tabular}
\end{table}

\section*{Acknowledgments}
	The author is grateful to Daniele Durante for his helpful comments, and acknowledges the support from
	MIUR-PRIN 2017 project 20177BRJXS.

\end{document}